\documentclass[11pt,a4paper,superscriptaddress,tightenlines,prfluids]{revtex4-2}
\usepackage{amssymb,amsmath}
\usepackage{ifthen,ifpdf}
\usepackage{verbatim}
\usepackage{graphicx,color}
\usepackage{mathtools}

\begin{document}
	
	\title{Chiral Edge Modes in Helmholtz-Onsager Vortex Systems}
	\author{Vishal P. Patil}
	\author{J\"orn Dunkel} 
	\affiliation{Department of Mathematics, Massachusetts Institute of Technology, 77 Massachusetts Avenue, Cambridge,~MA~02139, USA}
	\date{\today}

\begin{abstract}

Vortices play a fundamental role in the physics of two-dimensional (2D) fluids across a range of length scales, from quantum superfluids to geophysical flows. Despite a history dating back to Helmholtz, point vortices in a 2D fluid continue to pose interesting theoretical problems, owing to their unusual statistical mechanics. Here, we show that the strongly interacting Helmholtz-Onsager vortex systems can form statistical edge modes at low energies, extending a previously identified analogy between vortex matter and quantum Hall systems. Through dynamical simulations, Monte-Carlo sampling and mean-field theory, we demonstrate that these edge modes are associated with the formation of dipoles of real and image vortices at boundaries. The edge modes are robust, persisting in nonconvex domains, and are in quantitative agreement with the mean-field predictions.

\end{abstract}
	\maketitle

\section{Introduction}

The Helmholtz-Onsager system of point vortices~\cite{helmholtz1867lxiii, onsager1949statistical} is a widely studied model for inviscid flows in classical~\cite{chavanis1996statistical,taylor2009interacting,yin2003alternative,krishnamurthy2020liouville}  and quantum~\cite{bogatskiy2019edge,wiegmann2014anomalous, wiegmann2013hydrodynamics, bradley2012energy, yu2016theory, yu2017emergent,griffin2020magnus} fluid dynamics. As  demonstrated by Onsager~\cite{eyink2006onsager}, the classical point vortex system can condense into superclusters, thus reproducing a key element of two-dimensional (2D) turbulence phenomenology. This system has since emerged as a reduced model for turbulence, describing phenomena from spin-up turbulence~\cite{taylor2009interacting} in the classical case, to energy cascades~\cite{bradley2012energy} and turbulence decay~\cite{billam2014onsager} in the quantum case. In addition, the quantized point vortex equations reproduce certain phenomenological aspects of the quantum Hall effect~\cite{wiegmann2013hydrodynamics}. Experimentally, the point vortex model has recently been validated in quantum fluids~\cite{valani2018einstein,gauthier2019giant,johnstone2019evolution}, where point vortices emerge as topological defects~\cite{gauthier2019giant,johnstone2019evolution}. More generally, point vortices have become a minimal model for a variety of 2D turbulence phenomena, from atmospheric dynamics~\cite{eyink2006onsager} to biochemical signaling in cell membranes~\cite{tan2020topological}.

A compelling property of the point vortex system is its anomalous statistical mechanics, in which the phase space is bounded~\cite{eyink2006onsager} and equivalent to the configuration space. Previous important theoretical work has focused on the statistical properties of the high-energy regime~\cite{chavanis1996classification,chavanis2012kinetic,montgomery1992relaxation,buhler2002statistical,pointin1976statistical,esler2013statistical,esler2015universal,taylor2009interacting,montgomery1974statistical} in flat space and curved space~\cite{turner2010vortices}. In contrast, the low-energy dynamics in the presence of boundaries has not yet been widely explored. In this regime, Onsager's physical picture of strong attractive forces between vortices of opposite signs indicates the possibility of edge modes~\cite{eyink2006onsager}. A particularly interesting question is whether boundary induced image vortices can give rise to stable vortex localization through the formation of such edge modes in convex and nonconvex domains. Additionally, the presence of such modes would supplement previously established mappings between the point vortex model and quantum Hall type systems~\cite{wiegmann2013hydrodynamics}.

Here we investigate the formation and dynamics of statistical edge modes in confined point vortex systems at low energy. Complementing previous studies of point vortex statistics at intermediate-to-high energies~\cite{buhler2002statistical,chavanis1996classification, venaille2009statistical,venaille2011solvable}, we examine the robustness of these edge modes in convex and nonconvex domains. Comparing Monte Carlo sampling, dynamical simulations and mean-field theory, we find vortex edge modes in disk and bean-shaped geometries, which survive at large vortex numbers. Strikingly, these edge modes are a robust, real-space phenomenon, despite the strongly interacting nature of the point vortex system. Furthermore, these modes persist in nonconvex domains, reminiscent of the edge mode signatures observed in topological systems.

\section{Helmholtz -Onsager model}
\subsection{Kirchhoff equations}
The point vortex model consists of a system of $N$ point vortices in a 2D incompressible, inviscid fluid with constant density. Each vortex has strength $\lambda_a$ and position $\mathbf{x}_a(t) = (x_a(t), y_a(t))$ in a domain $\Omega$, which corresponds to the following vorticity field
\begin{align*}
\omega = \sum_{a}\lambda_a\delta\left(\mathbf{x} - \mathbf{x}_a(t)\right)
\end{align*}
The flow field $\mathbf{u}(t,\mathbf{x})$ for this vortex  configuration is obtained from the streamfunction~$\psi(t,\mathbf{x})$ as~\cite{eyink2006onsager}
\begin{align*}
\mathbf{u} = \nabla \psi \times \mathbf{e}_z 
, \qquad\qquad
\nabla^2 \psi = -\omega
\end{align*}
The streamfunction can be written in terms of Green's functions
\begin{align*}
\psi = -\sum_{a=1}^N\lambda_a G(\mathbf{x}, \mathbf{x}_a(t))
\end{align*}
where the Green's function~$G$ satisfies
\begin{align*}
\nabla^2 G(\mathbf{x},\mathbf{y}) & =\delta(\mathbf{x} - \mathbf{y}) 
,\qquad\qquad 
G|_{\partial \Omega} = 0
\end{align*}
The boundary condition for the Green's function arises from the fact that the boundary of the domain must be a streamline, $\psi |_{\partial \Omega} = \text{const.}$ For simply connected domains, we can set this constant to 0. Using this streamfunction as an ansatz for the inviscid vorticity equation~\cite{helmholtz1867lxiii, kirchhoff1883vorlesungen,wiegmann2013hydrodynamics}
\[\frac{\partial \omega}{\partial t} + \mathbf{u} \cdot\nabla\omega = 0\] 
yields a Hamiltonian system where $x_a$ and $y_a$ are canonically conjugate
\begin{align}\label{hamiltonian}
&\lambda_a \dot{x}_a = \frac{\partial \mathcal{H}}{ \partial y_a}  
,\qquad\qquad 
\lambda_a \dot{y}_a = -\frac{ \partial \mathcal{H}}{ \partial x_a} 
\end{align}
The Hamiltonian is given by
\begin{align}
\mathcal{H} &= -\sum_{a<b} \lambda_a \lambda_b G(\mathbf{x}_a, \mathbf{x}_b) - \frac{1}{2} \sum_a \lambda_a^2 g(\mathbf{x}_a)
\end{align}
where 
\[g(\mathbf{x}_a) = \lim_{\mathbf{x} \rightarrow \mathbf{x}_a} \left( G(\mathbf{x}, \mathbf{x}_a) - \frac{1}{2\pi} \log|\mathbf{x} - \mathbf{x}_a| \right)\]
is the regularized Green's function. The Hamiltonian coincides with the regularized kinetic energy of the fluid~\cite{menon2018statistical}
\begin{align*}
\mathcal{K} 
= \frac{1}{2} \int_\Omega d^2\mathbf{x} \,|\mathbf{u}|^2
= \frac{1}{2} \int_\Omega d^2\mathbf{x} \,\nabla\psi \cdot \nabla\psi
= \frac{1}{2} \int_\Omega d^2\mathbf{x} \, \psi \omega = \mathcal{H}
\end{align*}  
The quantities defined only depend on length $L$ and time $T$: $[\lambda] = L^2/T , \; [\mathcal{H}] = L^4/T^2$. Choosing a reference length scale $L_0$, we can define a timescale $T_0 = L_0^2/(N|\lambda|)$. Since the fluid density plays no role in the system, this also gives energy scale $E_0 = L_0^4/T_0^2$ and angular momentum scale $J_0 = L_0^4/T_0$. In particular
\[
\frac{E_0}{N ^2} = |\lambda|^2, \qquad\qquad \frac{J_0}{NL_0^2} = |\lambda|
\]
In the remainder, we consider systems of identical vortices with positive circulation, 
$$\lambda_a \equiv \lambda > 0$$ 
We choose units so that $L_0 = 1$ and $\lambda = 1$. This fixes the timescale, $T_0 = 1/N$, reflecting the fact that the fluid rotates faster as the number of point vortices increases. Below, we will focus on two different domains, the unit disk $D$ with radius $L_0=1$ and a bean-shaped domain $B$ obtained from~$D$ by a conformal mapping specified below.

\subsection{Disk-shaped domain}

\begin{figure*}[ht]
	\centering
	\includegraphics[width=\columnwidth]{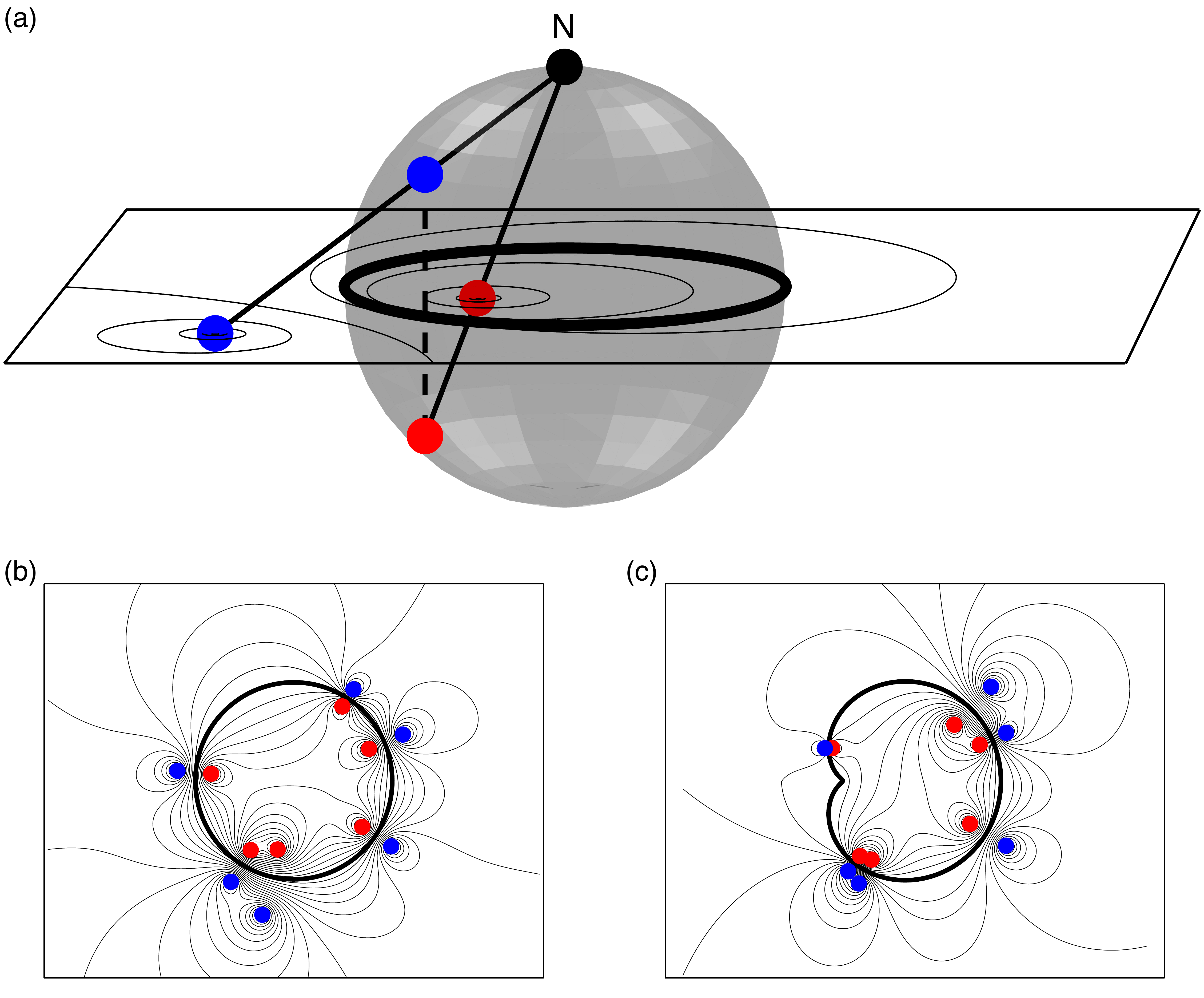}
	\caption{Image vortices drive boundary accumulation in symmetric and nonconvex domains.
		(a)~A real vortex with positive circulation inside the disk (red), gives rise to an image vortex with negative circulation outside (blue). The position of the image can be visualized through stereographic projection.
		(b)~At sufficiently low energies, image dipoles form at the boundary of the disk. Streamlines demonstrate the corresponding flow patterns.
		(c)~Image vortex dipoles persist upon conformal mapping to a bean-shaped domain.
	} 
	\label{fig_image}
\end{figure*}

In complex coordinates, $z = x+iy$, the Green's functions $G$ and $g$ for the disk $D$ are~\cite{lin1941motion1,lin1941motion2,crowdy2010new}
\begin{align*}
G^D(z,w) &= \frac{1}{2\pi} \log \left|\frac{z-w}{z^* w -1 }\right| \qquad\qquad g^D(z) = -\frac{1}{2\pi}\log|1-z^*z|
\end{align*}
This gives a compact expression for the Hamiltonian
\begin{align*}
\mathcal{H} &= -\frac{1}{2\pi}\sum_{a<b} \lambda^2 \log \left|\frac{z_a-z_b}{z_a^* z_b -1 }\right| + \frac{1}{4\pi} \sum_a \lambda^2 \log|1-z_a^*z_a|\\
\end{align*}
and the velocity
\begin{align*}
\lambda\dot{z}_a  = -2i\frac{\partial \mathcal{H}}{\partial \bar{z}_a} &= \frac{i\lambda^2}{2\pi} \sum_{b\neq a} \frac{1}{z_a^* - z_b^*} - \frac{i\lambda^2}{2\pi} \sum_{b} \frac{1}{z_a^* - (1/z_b)}
\end{align*}
The first sum captures the flow generated by point vortices at the locations $z_b$, in the the absence of boundaries. Analogously, the second sum, which includes $b=a$, is due to image vortices located at the points $1/z_b^*$ [Fig.~\ref{fig_image}(a)]. The velocity of the whole fluid, $\mathbf{u}(t,z) = (u(t,z),v(t,z))$, is
\begin{align*}
u(t,z)+iv(t,z) = \frac{i\lambda}{2\pi} \sum_{a} \left[\frac{1}{z^* - z_a^*} -  \frac{1}{z^* - (1/z_a)}\right]
\end{align*}
The Hamiltonian for the unit disk can be split up into a bulk energy and a boundary energy, arising from the interaction between real vortices and image vortices [Figs.~\ref{fig_image}(a) and \ref{fig_image}(b)]
\begin{align*}
\mathcal{H} &= -\frac{1}{4\pi}\sum_{a\neq b} \lambda^2 \log \left|z_a-z_b\right| + \frac{1}{4\pi} \sum_{a,b} \lambda^2 \log|1-z_a z_b^*|
\end{align*}
The singularities of $\mathcal{H}$ correspond to the locations of these real and image vortices. The energy is singular when the $a$'th vortex approaches the real vortex at $z_b$ or the image vortex at $1/z_b^*$ [Figs.~\ref{fig_image}(a) and \ref{fig_image}(b)].

\subsection{Nonconvex bean-shaped domain}

 A bean-shaped domain $B$ can be obtained from the unit disk by a conformal mapping. Consider the image of $D$ under a map of the form $f^{-1}(z) = (b'z-1)^2/a'$. When $b'=1$, this mapping produces a nonconvex cardioid domain, which posseses a cusp singularity at the boundary. By taking $b'=0.9$, we instead obtain a smooth nonconvex bean-shaped domain, $B$ [Fig.~\ref{fig_image}(c)]. We set $a'^2 = 2b'^2(b'^2 + 2)$ so that $B$ has area $\pi$, equal to the unit disk. The Green's function in $B$ transforms with the conformal map $f$
\begin{align*}
G^B(z,w) &= G^D(f(z), f(w)) = \frac{1}{2\pi} \log \left|\frac{f(z)-f(w)}{f(z)^* f(w) -1 }\right|
\end{align*}
and the regularized Green's function is
\begin{align*}
g^B(z) &= \lim_{w\rightarrow z} \left( \frac{1}{2\pi} \log \left| \frac{f(z) - f(w)} { f(z)^*f(w) -1} \right| - \frac{1}{2\pi} \log |z - w|\right)\\
&= -\frac{1}{2\pi}\left( \log|1 - f(z)^*f(z)| - \log | f'(z)|\right)
\end{align*}
As before, the form of the Hamiltonian indicates the presence of image vortices
\begin{align}\label{H}
\mathcal{H} &= -\frac{1}{4\pi}\sum_{a\neq b} \lambda^2 \log \left|f(z_a)-f(z_b)\right| + \frac{1}{4\pi} \sum_{a,b} \lambda^2 \log|1-f(z_a) f(z_b)^*| - \frac{1}{4\pi} \sum_{a} \lambda^2 \log|f'(z_a)|
\end{align}
The positions of real and image vortices again follow from the singularities of the Hamiltonian. The first term is singular when $z_a$ approaches the real vortex at $z_b$, and the second term is singular when $z_a$ approaches the image vortex at $f^{-1}\left(1/f(z_b)^*\right)$ [Fig.~\ref{fig_image}(c)]. Furthermore, the first and third terms of the Hamiltonian in (\ref{H}) are bounded below, unlike the second term. To see this, observe that the boundedness of the domain means that the expressions $-\log|f(z_a) - f(z_b)|$ in the first term are bounded below. Similarly, $-\log|f'(z_a)|$ is bounded below whenever $f'(z_a)$ is bounded, as is the case for the mappings we consider. On the other hand, the second term of (\ref{H}) approaches negative infinity when real vortices approach image vortices, although this divergence is not independent of divergences in the first term. Nevertheless, this heuristic argument indicates that an effective image vortex attraction at low energies will give rise to vortex clustering at the boundary of the domain. This image vortex dipole mechanism (Fig.~\ref{fig_image}) is responsible for the formation of edge modes.

\subsection{Conserved quantities}

In addition to the energy $\mathcal{H}$, an incompressible, inviscid fluid in 2D also has conservation laws for linear momentum $\mathcal{P}$, angular momentum $\mathcal{A}$, vorticity $\mathcal{V}$, and enstrophy $\mathcal{E}$ 
\begin{align*}
\mathcal{P} = \int_\Omega d^2\mathbf{x} \, \mathbf{u} , \qquad 
\mathcal{A} = \int_\Omega d^2\mathbf{x} \, \left(\mathbf{x} \wedge \mathbf{u} \right) ,\qquad
\mathcal{V} = \int_\Omega d^2\mathbf{x} \, \omega,\qquad 
\mathcal{E} = \int_\Omega d^2\mathbf{x} \, \omega^2
\end{align*} 
where the pseudoscalar $\mathbf{x} \wedge \mathbf{u}= xv-yu$ corresponds to the $z$-component of the 3D  cross-product. The linear momentum vanishes in a bounded domain, which can be shown componentwise using the streamfunction $\mathbf{u} = (\psi_y, -\psi_x)$
\begin{align*}
\mathcal{P} \cdot \mathbf{e}_x = \int_\Omega d^2\mathbf{x} \, \nabla \cdot \left(\psi\mathbf{e}_y\right) = \int_{\partial \Omega} d\mathbf{n} \cdot \mathbf{e}_y \psi = 0
\end{align*}
where $d\mathbf{n}$ is the normal curve element. The second equality follows from the divergence theorem and the final equality uses the fact that $\psi |_{\partial \Omega}$ is constant and $\partial \Omega$ is a closed curve.
The vorticity and enstrophy have straightforward expressions in terms of the vortex strengths, however the enstrophy is singular
\begin{align*} 
\mathcal{V} = \sum_a \lambda_a = N\lambda ,\qquad\qquad 
\mathcal{E} = \sum_{a,b} \lambda_a\lambda_b \delta(\mathbf{x}_a - \mathbf{x}_b) = \lambda^2\sum_{a,b}\delta(\mathbf{x}_a - \mathbf{x}_b) 
\end{align*} 
In the systems of identical point vortices we consider here, the conservation of $\mathcal{V}$ is equivalent to the conservation of particle number, $\mathcal{V} = N$. The conversation law for angular momentum depends on the pressure field $p(t,\mathbf{x})$, which can be found in terms of the velocity field, $\mathbf{u}$, through the Euler equation for an inviscid fluid
\begin{align*}
-\nabla p  = \frac{\partial \mathbf{u}} {\partial t} + \mathbf{u} \cdot \nabla\mathbf{u}
\end{align*}
Taking the cross product of the Euler equation with $\mathbf{x}$ gives an equation for angular momentum transport; integrating this transport equation and using the divergence theorem yields the conservation law for angular momentum in terms of the boundary advection of angular momentum and the boundary torque due to pressure
\begin{align*}
\frac{d}{dt} \mathcal{A} &= -\int_{\partial \Omega} d\mathbf{n}\cdot \mathbf{u}\, \left(\mathbf{x} \wedge \mathbf{u}\right) + \int_{\partial \Omega} d\mathbf{n} \wedge \mathbf{x} \,p = \int_{\partial \Omega} d\mathbf{n} \wedge \mathbf{x} \,p
\end{align*}
The advection term vanishes since $\mathbf{u}$ has no normal component on the boundary. Angular momentum is globally conserved if the boundary torque vanishes, which occurs when the domain $\Omega$ has the appropriate symmetry~\cite{eyink2006onsager,pointin1976statistical}. Using Stokes' theorem, the angular momentum of the fluid can be expressed as 
\begin{align*}
\mathcal{A} = \frac{1}{2} \int_{\Omega} d^2\mathbf{x} \, \left[ \nabla \wedge \left(\mathbf{u}|\mathbf{x}|^2\right) -\omega|\mathbf{x}|^2\right]  = \frac{1}{2}\int_{\partial \Omega} d\mathbf{r} \cdot \mathbf{u}\,|\mathbf{x}|^2  - \frac{1}{2}\mathcal{J}
\end{align*}
where $d\mathbf{r}$ is the tangential curve element and
\[\mathcal{J} = \sum_a \lambda |\mathbf{x}_a|^2\]
If the domain has circular symmetry, $\mathcal{A}$ is conserved. In particular, when $\Omega$ is the unit disk $D$, the circulation term can be simplified
\begin{align*}
\mathcal{A} = \frac{1}{2} \sum_a \lambda_a - \frac{1}{2} \sum_a \lambda_a |\mathbf{x}_a|^2 = \frac{1}{2}\mathcal{V} - \frac{1}{2} \mathcal{J}
\end{align*}
where $\mathcal{J}$ is the conserved vortex angular momentum. The statistical mechanics of the system can be described in terms of the nonzero and nonsingular conserved quantities $\mathcal{H}, \mathcal{A}$ and $\mathcal{V}$.

\subsection{Density of states}

\begin{figure*}[ht]
	\centering
	\includegraphics[width=\columnwidth]{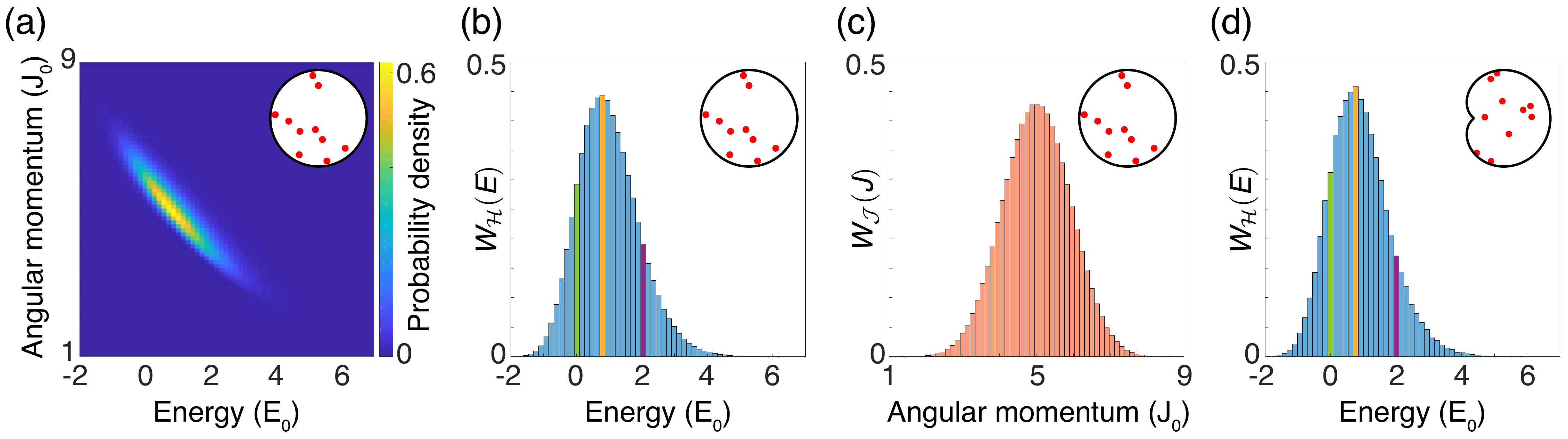}
	\caption{Density of states for the unit disk and the bean of area $\pi$ for $N=10$ vortices.
		(a)~The joint density of states over energy, $\mathcal{H}$, and angular momentum, $\mathcal{J}$, in the disk reveals a strong correlation between $\mathcal{H}$ and $\mathcal{J}$.
		(b)~The marginal density of states over energy exhibits distinct low-energy (green) and high-energy (purple) regimes on either side of the mode, $E^*$, of the distribution (yellow).
		(c)~The marginal density of states over angular momentum is symmetric.
		(d)~The density of states in the bean-shaped domain exhibits the same pattern of low- and high-energy regimes as the disk.} 
	\label{fig_density_of_states}
\end{figure*}

The different statistical regimes of the point vortex system can be identified through the density of states. For the disk $D$, conservation of both $\mathcal{H}$ and $\mathcal{J}$ implies there is a joint density of states~$W(E,J)$ [Fig.~\ref{fig_density_of_states}(a)]. Defining $\xi = (x_1,y_1, x_2, y_2, ..., x_N, y_N)$, we have
\begin{align*}
W(E,J) &= \frac{1}{\pi^N} \int_{D^N} d \xi \, 
\delta\left(\mathcal{H}(\xi) - E\right)
\delta\left(\mathcal{J}(\xi) - J\right)
\end{align*}
The marginal densities $W_\mathcal{H}(E)$ and $W_\mathcal{J}(J)$ over energy and angular momentum follow from the joint density by [Fig.~\ref{fig_density_of_states}(b) and \ref{fig_density_of_states}(c)]
\begin{align*}
W_\mathcal{H}(E) &= \frac{1}{\pi^N} \int_{D^N} d\xi\, \delta\left(\mathcal{H}(\xi) - E\right)\\
W_\mathcal{J}(J) &= \frac{1}{\pi^N} \int_{D^N} d\xi\, \delta\left(\mathcal{J}(\xi) - J\right)
\end{align*}
Flows where $\mathcal{J}$ is large in magnitude must involve a concentration of vortices at the boundary, and thus $\mathcal{J}$-conservation may be thought of as an angular momentum barrier. However, due to the correlation between $\mathcal{H}$ and $\mathcal{J}$ [Fig.~\ref{fig_image}(a)], it is hard to disentangle this effect from the low-energy boundary attraction described above.

As is typical of a generic bounded domain, the bean-shaped domain has no additional conserved quantities, so there is only a density of states over energy [Fig.~\ref{fig_density_of_states}(d)]. This density, $W_\mathcal{H}(E)$, has a maximum at some energy, $E=E^*$, which arises from the fact that the phase space is bounded [Figs.~\ref{fig_density_of_states}(b) and \ref{fig_density_of_states}(d)]. Here, we focus on the low-energy regime, $E<E^*$.

\section{Monte Carlo simulations and Hamiltonian dynamics predict edge modes}

\begin{figure*}[ht]
	\centering
	\includegraphics[width=\columnwidth]{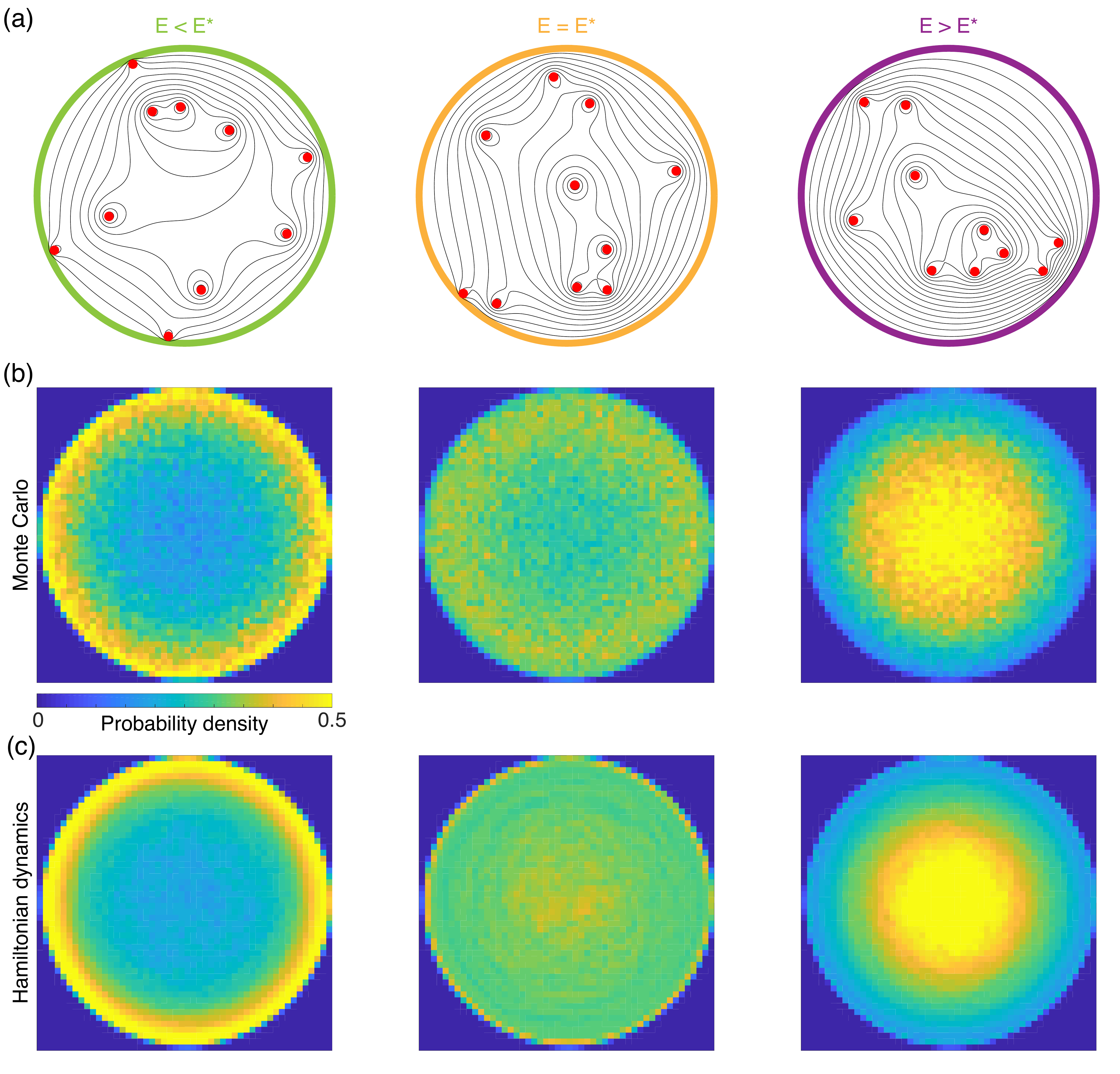}
	\caption{Monte Carlo sampling for the disk agrees with time-averaged Hamiltonian dynamics at low, 
		intermediate and high energies ($N=10$).
		(a)~Streamlines demonstrate the flow patterns of sample points in phase space with given energy.
		(b)~Monte Carlo sampling at fixed energy, and varying angular momentum reveals low-energy edge modes and high-energy vortex clustering in the disk; sample size $>10^4$ configurations. 
		(c)~The time-averaged dynamics of the vortex system agrees with the Monte Carlo predicted statistical states. Energy from left to right in (a),(b),(c): $E=0.0,0.8,2.0$. Angular momentum from left to right in (c): $J=5.8,4.9,4.0$. Simulation time: $T = 10^5 T_0$.} 
	\label{fig_disk_1}
\end{figure*}

\begin{figure*}[ht]
	\centering
	\includegraphics[width=\columnwidth]{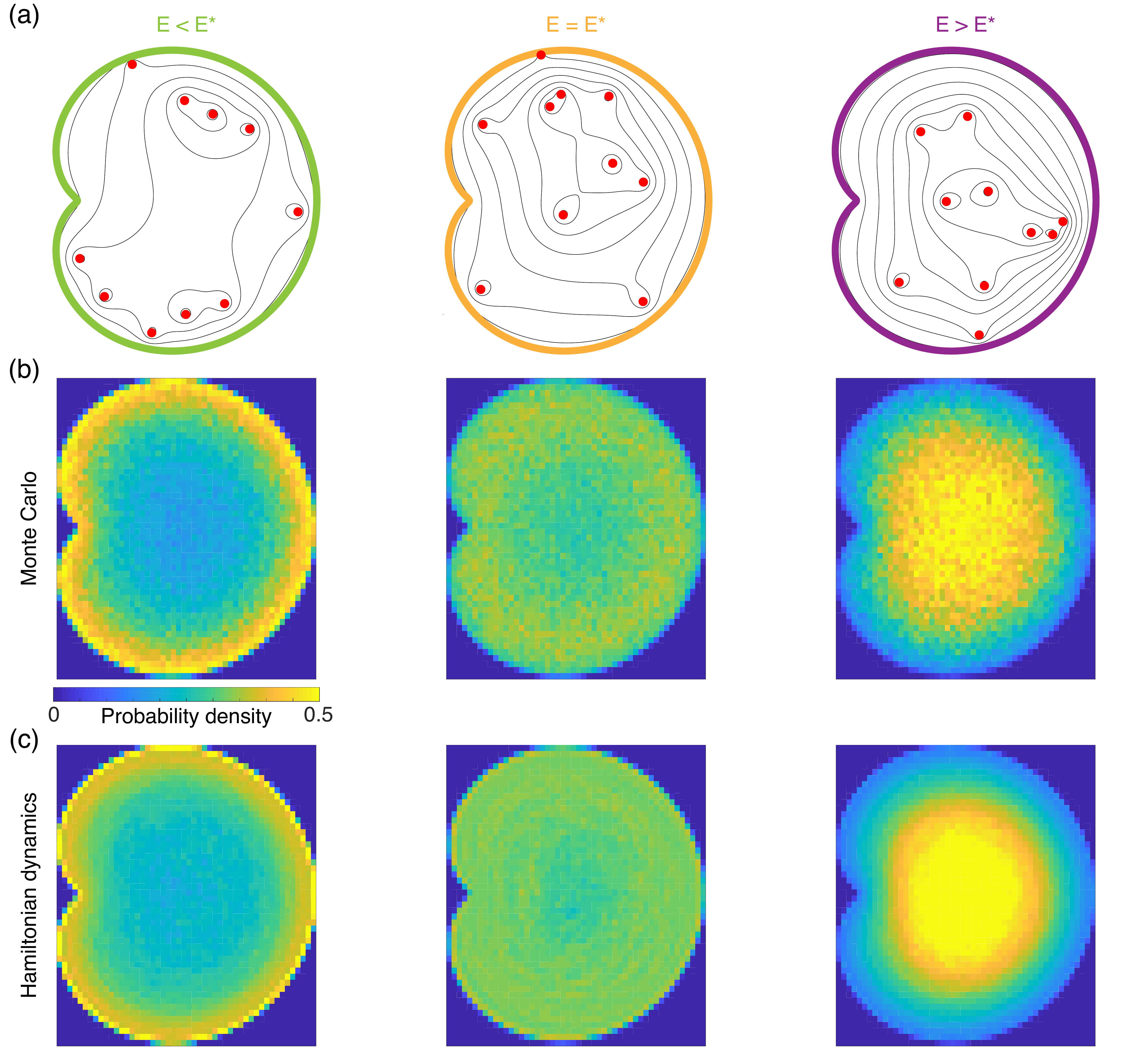}
	\caption{Monte Carlo sampling for the bean also agrees with time-averaged Hamiltonian dynamics at low, 
		intermediate and high energies ($N=10$).
		(a)~Streamlines demonstrate the flow patterns of sample points in phase space with given energy.
		(b)~Monte Carlo sampling reveals low-energy edge modes and high-energy clustering in the bean; sample size $>10^4$ configurations.
		(c)~The time-averaged dynamics of the vortex system agrees with the Monte Carlo predicted statistical states. Energy from left to right [(a),(b),(c)]: $E=0.0,0.8,2.0$. Simulation time: $T = 10^5 T_0$.} 
	\label{fig_bean_1}
\end{figure*}

Boundary attraction through image vortices leads to the formation of a low-energy statistical edge mode in both the disk (Fig.~\ref{fig_disk_1}) and the bean (Fig.~\ref{fig_bean_1}) at small vortex number. As discussed above, the Hamiltonian for a general domain, obtained from the image of the unit disk under a conformal mapping $f^{-1}$, is
\begin{align*}
\mathcal{H} &= -\frac{1}{4\pi}\sum_{a\neq b} \lambda^2 \log \left|f(z_a)-f(z_b)\right| + \frac{1}{4\pi} \sum_{a,b} \lambda^2 \log|1-f(z_a) f(z_b)^*| - \frac{1}{4\pi} \sum_{a} \lambda^2 \log|f'(z_a)|
\end{align*}
Only the second term is unbounded below, so states with sufficiently low energy are typically only produced when $f(z_a)f(z_b)^*$ is close to 1 for enough pairs of vortices, or equivalently, when enough vortices are close to the boundary. For the disk, such a configuration will have large angular momentum, $\mathcal{J} = \sum_a \lambda |\mathbf{x}_a|^2$, which explains the $\mathcal{H}-\mathcal{J}$ correlation [Fig.~\ref{fig_density_of_states}(a)]. The agreement between the distributions produced by time-averaged Hamiltonian dynamics and Monte Carlo sampling (Figs.~\ref{fig_disk_1} and \ref{fig_bean_1}) also suggests that a mean-field approach can be used to formalize the above heuristic argument for edge modes.

\section{Mean-field limit}
\subsection{Edge modes persist in the mean-field limit}
To identify the low-energy regime for large $N$ and understand the corresponding statistical states, we return to the density of states for a general bounded domain
\begin{align*}
W_\mathcal{H}(E) &= \frac{1}{\pi^N} \int_{\Omega^N} d\xi\, \delta\left(\mathcal{H}(\xi) - E\right)
\end{align*}
The function $W_\mathcal{H}(E)$ can also be understood as the probability density function (pdf) of the energy when $N$ points are chosen uniformly at random in the domain. Let $Z_a = X_a + iY_a$ be a collection of such uniform random variables for $a = 1$ to $N$. The energy is now a random variable $\hat{H}$
\begin{align*}
\hat{H} = -\frac{1}{4\pi} \sum_{a\neq b}  G(Z_a, Z_b) + \frac{1}{4\pi}\sum_a g(Z_a)
\end{align*}
where the first sum contains $N(N-1)$ terms, the second sum contains $N$ terms and the strength of each vortex $\lambda$ has been set to 1 as before. Previous work~\cite{esler2017statistics} has identified the limiting distribution of $\hat{H}$ in bounded domains. For the identical vortex systems considered here, this limiting distribution is Gaussian. An informal derivation of this fact follows from the theory of $U$-statistics~\cite{hoeffding1992class,o1991limiting}. Set $\bar{H} = \hat{H}/N^2$, and consider  
\begin{equation}
\label{density_limit}
\sqrt{N}(\bar{H}-\mu) = \sqrt{N}\left(-\frac{1}{4\pi N^2} \sum_{a\neq b}  G(Z_a, Z_b) - \mu\right) + \frac{1}{4\pi N\sqrt{N}}\sum_a g(Z_a)
\end{equation}
where
$$
\mu = -\frac{1}{4\pi}\mathbb{E} \left[G(Z_a, Z_b)\right]
$$
The second term in \eqref{density_limit} is a sum of only $N$ independent random variables, and thus converges to $0$. A central limit type theorem~\cite{hoeffding1992class} for the sequence of $N(N-1)$ random variables $G(Z_a, Z_b)$ shows that $\sqrt{N}(\bar{H} - \mu)$ is asymptotically normal, where $\mu$ and the variance $\sigma^2$ are given by~\cite{hoeffding1992class,campbell1991statistics,esler2017statistics}
\begin{align*}
\mu = -\frac{1}{4\pi}\mathbb{E} G(Z_a, Z_b) , \qquad\qquad \sigma^2 =\lim_{N\rightarrow \infty} N\,\text{Var}(\bar{H}) = \lim_{N\rightarrow \infty} \frac{\text{Var}(\hat{H})}{N^3}
\end{align*}
Furthermore, $U$-statistics theory~\cite{hoeffding1992class} indicates that $\sigma^2>0$ holds whenever $\mathbb{E}(G(z,Z_a))$ is not constant as a function of $z$. Formalizing the above argument would require showing that $G$ and $g$ meet suitable regularity conditions~\cite{hoeffding1992class}. The scaling of the variance is $\text{Var}(\bar{H}) \sim O(1/N)$, analogous to the central limit theorem. This scaling may be understood intuitively as a consequence of the fact that there are only $N$, and not $N^2$, independent points. Although $\mathbb{E}(\hat{H}/N^2) \sim O(1)$, as $N\rightarrow \infty$, there is an $O(1/N)$ contribution to the mean coming from the boundary terms, $\mathbb{E}(g(Z_a))$. For finite $N$, the pdf of $\hat{H}/N^2$ can therefore be better approximated by a Gaussian with mean
\[
\mathbb{E}\left(\frac{\hat{H}}{N^2} \right)= \mu +\frac{\mu_b}{N}
\]
where $\mu_b = \mathbb{E}(g(Z_a))/4\pi$. The limit allows us to define the low-energy regime in terms of a rescaled energy $\epsilon$, which measures the distance from the mean energy in units of standard deviation
\begin{align}
\epsilon = \frac{\sqrt{N}}{\sigma}\left[\frac{\mathcal{H}}{N^2} - \left(\mu + \frac{\mu_b}{N}\right)\right]
\end{align} 
Similarly, for the disk, the angular momentum becomes a random variable $\hat{J}$
\begin{align*}
\hat{J} = \sum_a |Z_a|^2
\end{align*}
where $\mathbb{E}(|Z|^2) = 1/2$ and $\text{Var}(|Z|^2) = 1/12$. Using the central limit theorem, the appropriately rescaled angular momentum is therefore
\begin{align*}
\ell = \sqrt{12N}\left(\frac{\hat{J}}{N} - \frac{1}{2}\right)
\end{align*}

\subsection{Mean-field theory in the disk}

\begin{figure*}[b]
	\centering
	\includegraphics[width=\columnwidth]{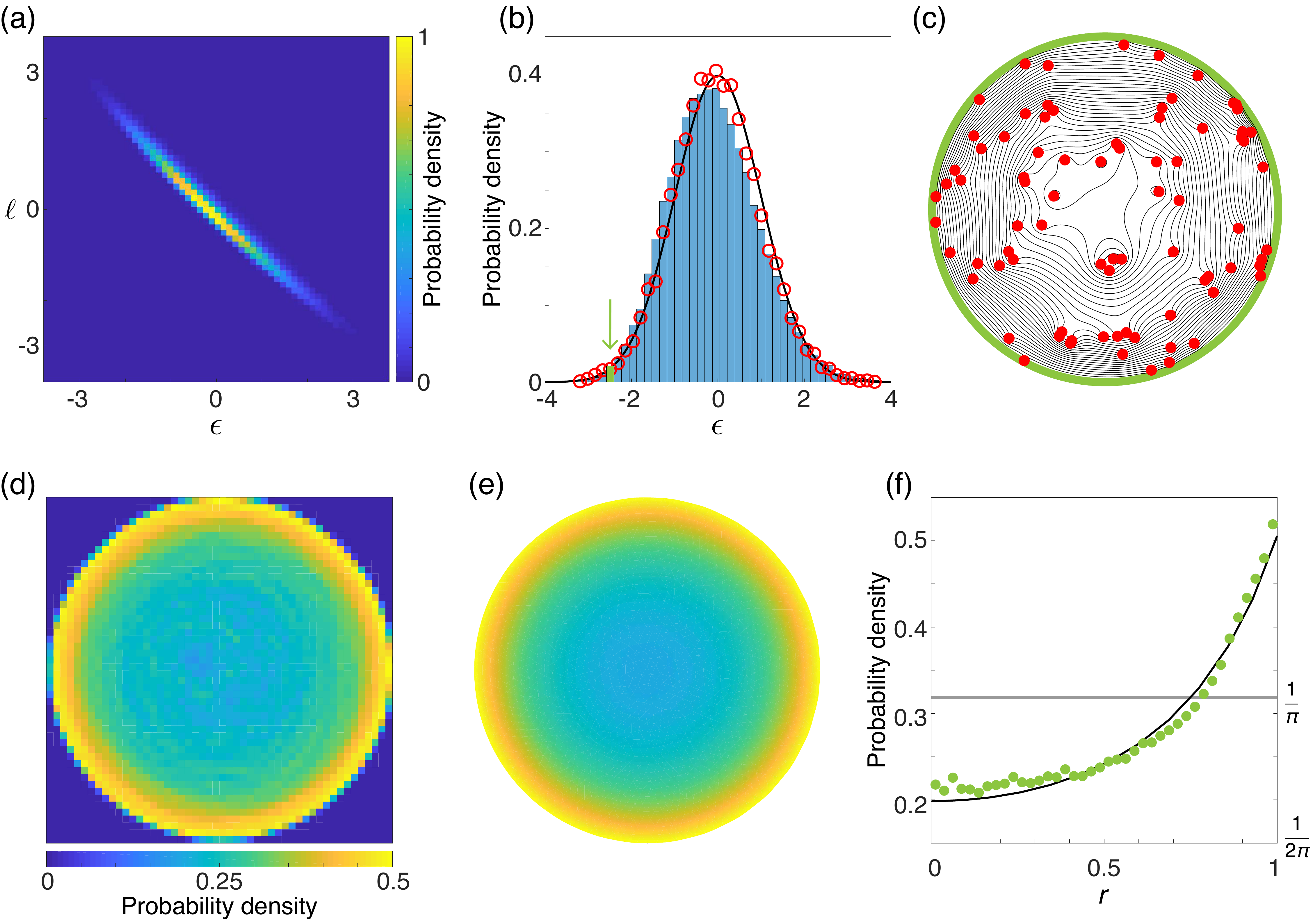}
	\caption{Low-energy edge modes persist at large vortex numbers and are in quantitative agreement with mean-field predictions. ($N=80$).
		(a)~Rescaled energy ($\epsilon$) and angular momentum ($\ell$) are strongly correlated at large $N$.
		(b)~The density of states over $\epsilon$ for $N=80$ (blue histogram) and $N=1000$ (red circles) demonstrates the convergence to the limiting density (solid curve). The green line selects the rescaled energy $\epsilon = -2.5$ at $N=80$.
		(c)~Streamlines demonstrate the flow pattern of a sample point in phase space with low energy, $\epsilon = -2.5$ (see movie S1).
		(d)~The time-averaged Hamiltonian dynamics of $N=80$ vortices at $\epsilon = -2.5$ exhibits edge modes (see movie S2). Simulation time: $T = 8\times 10^3 T_0$.
		(e)~Mean-field theory in the disk predicts edge modes ($\beta = 15, \gamma=0$).
		(f)~Predicted and measured probability density as a function of the radial coordinate $r$. The zero-angular momentum mean-field prediction (solid black curve) agrees quantitatively with time-averaged Hamiltonian dynamics ($N=80$, green circles), and deviates significantly from the uniform vortex distribution (solid horizontal line).
		} 
	\label{fig_disk_mf}
\end{figure*}

\begin{figure*}[ht]
	\centering
	\includegraphics[width=\columnwidth]{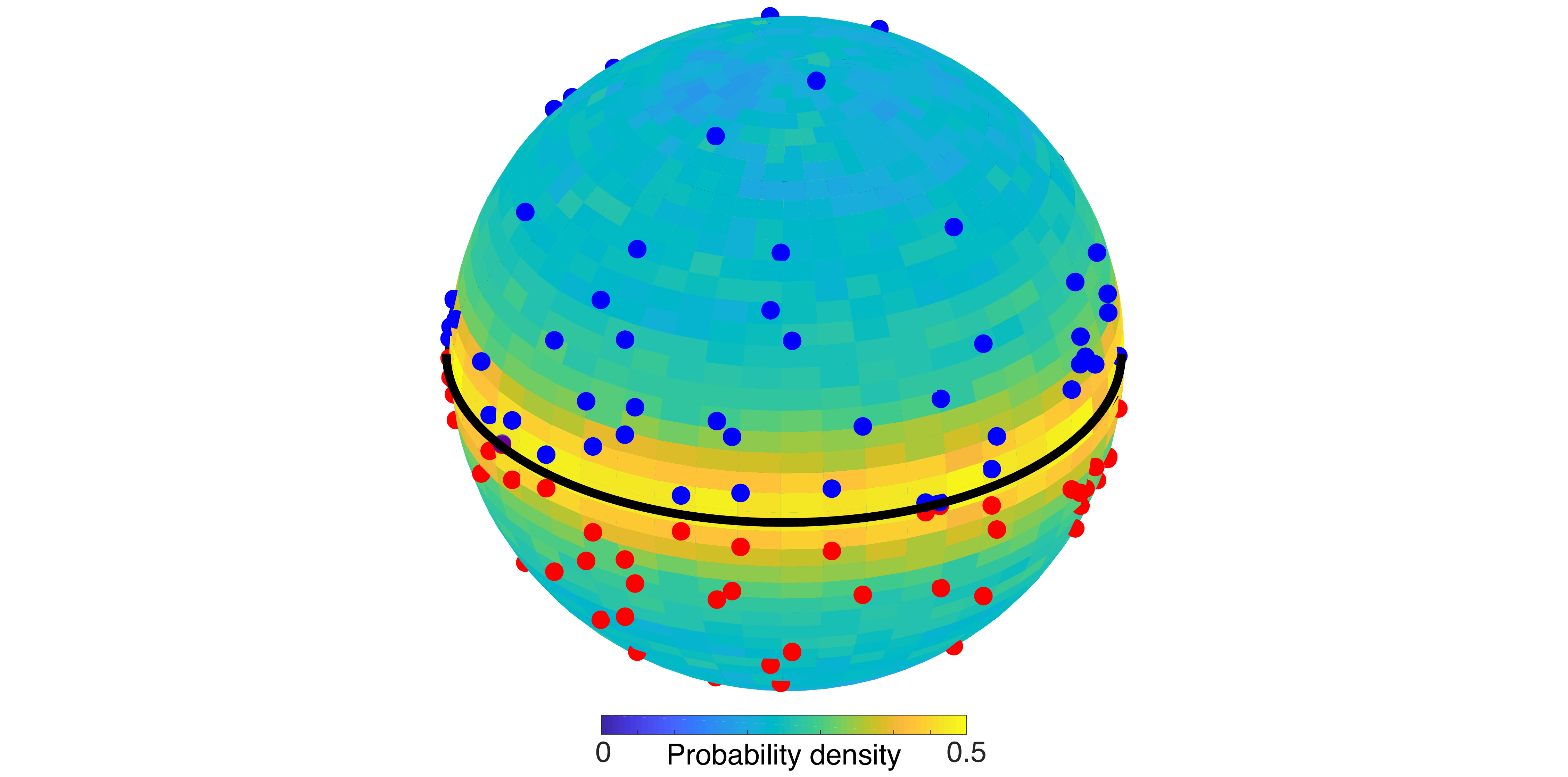}
	\caption{Visualization of edge modes in the unit disk from Fig.~\ref{fig_disk_mf}(d) through stereographic projection ($N=80$). The southern hemisphere, containing positive vortices (red) has $\lambda=+1$, and the northern hemisphere, containing negative image vortices (blue) has $\lambda=-1$. The vortex circulation $\lambda$ changes sign across the equator, leading to the formation of an edge mode.} 
	\label{fig_sphere_density}
\end{figure*}

Statistical edge modes persist in the disk at low energy for large vortex numbers $N$. When $N$ is sufficiently large, the densities of state approach their limiting values [Figs.~\ref{fig_disk_mf}(a) and \ref{fig_disk_mf}(b)], and the rescaled energy $\epsilon$ defines the low-energy regime. The robustness of edge modes at low $\epsilon$ [Figs.~\ref{fig_disk_mf}(c) and \ref{fig_disk_mf}(d)], can be understood analytically through the mean-field approach of maximizing the entropy of the vortex system at fixed energy, particle number, and angular momentum if the domain has circular symmetry~\cite{miller1990statistical,robert1991maximum,menon2018statistical}. It is convenient to explicitly state factors of the vortex circulation $\lambda$ in the mean-field picture. We set $\rho = \omega L_0^2 / (N\lambda) $ to be the dimensionless vortex density, so $\rho$ integrates to $L_0^2 =1$. Similarly, we rescale the streamfunction, $\tilde{\psi} = \psi L_0^2/N$ so that $\nabla^2\tilde{\psi} = -\lambda\rho$. The constrained entropy functional is~\cite{chavanis2012kinetic}
\begin{align}\label{MFT_functional}
\mathcal{I}[\rho] &= S -  \beta \lambda^2 T_0^2 \mathcal{H} - \alpha \int_\Omega d^2\mathbf{x} \, \rho  -  \frac{\beta\gamma L_0^2}{N} T_0 \mathcal{J} \\
&=  -\int_\Omega d^2\mathbf{x} \, \rho \log \rho - \beta\lambda \int_\Omega d^2\mathbf{x} \, \tilde{\psi} \rho
  -  \alpha\int_\Omega d^2\mathbf{x} \, \rho  -  \beta\gamma\lambda \int_\Omega d^2\mathbf{x} \, r^2 \rho \nonumber
\end{align}
where $\beta$, $\alpha$, $\gamma$ are Lagrange multipliers enforcing the constraints of constant energy, particle number, and angular momentum, respectively, with units $[\beta] = T^2/L^6$, $[\gamma]=L^2/T, [\alpha] = 0$. If $\Omega$ does not have circular symmetry, then $\gamma=0$. By analogy with the first law of thermodynamics, $\beta$ and $\alpha$ are sometimes thought of as an inverse temperature and a chemical potential. Extremizing $\mathcal{I}$ yields
\begin{align*}
\frac{\delta \mathcal{I}}{\delta \rho} &= -\log\rho - 1 - \beta\lambda\tilde{\psi} - \beta\lambda\gamma r^2 - \alpha = 0
\end{align*}
and thus
\begin{align*}
\rho = e^{-1-\alpha} e^{-\beta\lambda\left(\tilde{\psi}+\gamma r^2\right)} = \frac{1}{Z} e^{-\beta\lambda\left(\tilde{\psi}+\gamma r^2\right)}
\end{align*}
where $Z = e^{1+\alpha}$.
Using the streamfunction-vorticity relation, $\nabla^2\tilde{\psi} = - \lambda\rho$, gives the mean-field equation
\begin{align}\label{MFE}
\nabla^2\tilde{\psi} = 
-\frac{\lambda}{Z} e^{-\beta\lambda\left(\tilde{\psi}+\gamma r^2\right)} 
, \qquad\qquad 
\tilde{\psi}|_{\partial\Omega} = 0 
\end{align}
where
\begin{align}
Z = \int_\Omega d^2\mathbf{x} \, e^{-\beta\lambda(\tilde{\psi}+\gamma r^2)}
\end{align}
By setting $\phi = -\beta\lambda\tilde{\psi}$ and $\beta_1 = \beta/Z$, the mean-field equation can be expressed as a nonlinear eigenvalue problem
\begin{align*}
\nabla^2 \phi &= \beta_1 \lambda^2 e^{-\beta\gamma\lambda r^2} e^\phi , \qquad\qquad \phi|_{\partial \Omega} = 0
\end{align*}
where
\begin{align*}
\beta &= \beta_1\lambda^2 \int_\Omega d^2\mathbf{x} \, e^{-\beta\lambda\gamma r^2} e^\phi
\end{align*}
In the disk, we can simplify this equation by assuming axisymmetry, $\phi = \phi(r)$, and neglecting angular momentum, $\gamma=0$, based on the strong correlation between angular momentum and energy [Fig.~\ref{fig_disk_mf}(a)]
\begin{align*}
\phi'' +\frac{1}{r} \phi' - \beta_1\lambda^2 e^\phi = 0, \qquad \phi'(0) = 0, \quad \phi(1)=0
\end{align*}
The boundary condition at 0 comes from the requirement that $\phi$ is smooth at the origin. This equation can be solved exactly to give a vortex density $\rho$ with one free parameter~\cite{menon2018statistical}
\begin{align}\label{mf_sol}
\rho(r) = \frac{8}{8\pi+\beta} \frac{1}{\left(1 - \frac{\beta r^2}{8\pi+\beta}\right)^2} , \qquad \beta > -8\pi
\end{align}
This solution agrees quantitatively with simulations of Hamiltonian dynamics [Figs.~\ref{fig_disk_mf}(e) and \ref{fig_disk_mf}(f)], confirming the validity of the mean-field theory at low energies. Furthermore, the mean-field solution predicts a wide range of low-energy edge modes. Whenever $\beta>0$, the solution (\ref{mf_sol}) is maximized on the boundary. For large $\beta$, the solution describes an increasingly concentrated edge mode. The role of image vortices in this edge mode solution can be visualized by mapping the system onto a sphere [Figs.~\ref{fig_image} and \ref{fig_sphere_density}].

\begin{figure*}[ht]
	\centering
	\includegraphics[width=\columnwidth]{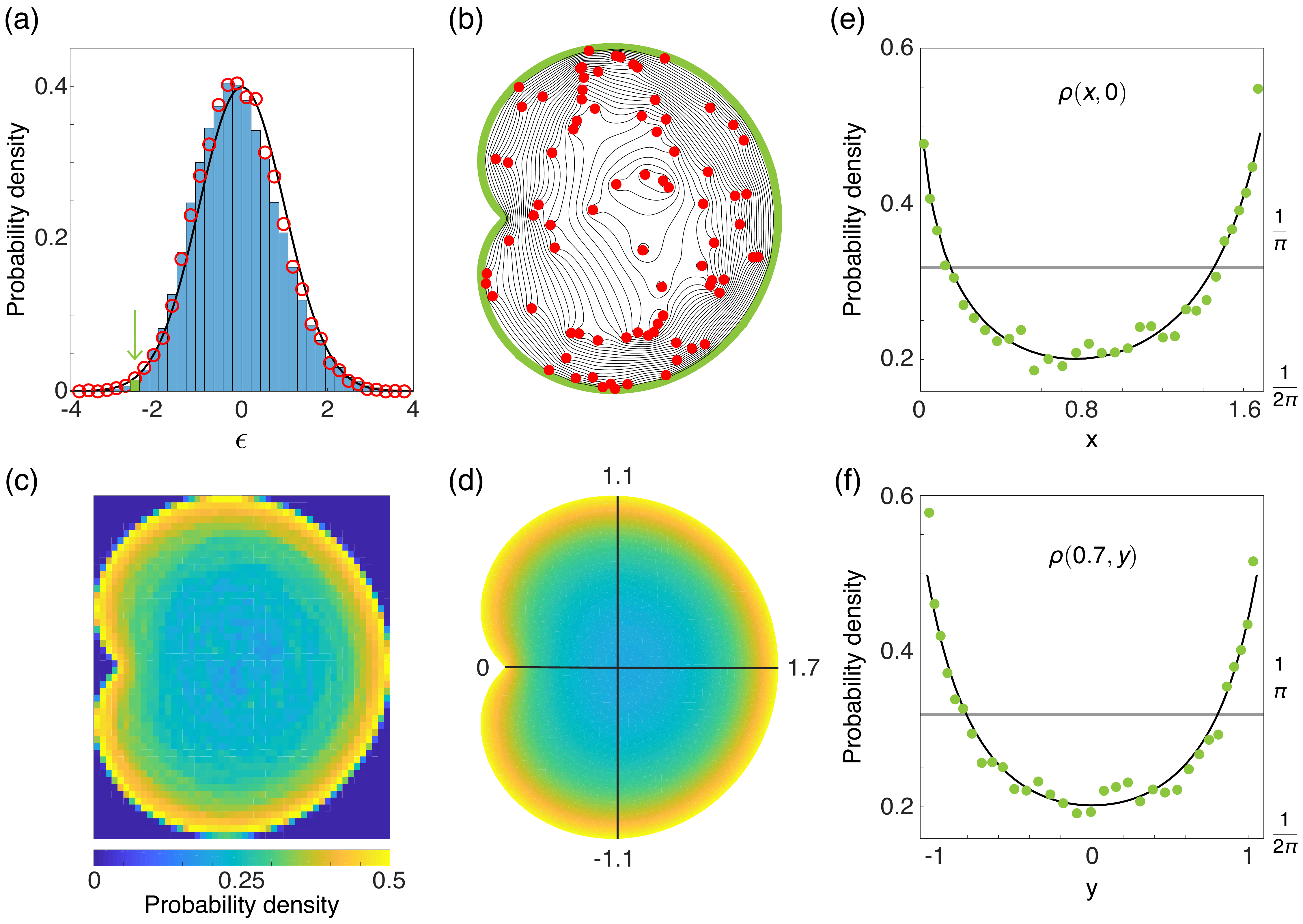}
	\caption{Low-energy edge modes persist at large vortex numbers and are in quantitative agreement with mean-field predictions.
		(a)~The density of states over rescaled energy ($\epsilon$) for $N=80$ (blue histogram) and $N=1000$ (red circles) demonstrates convergence to the limiting density (solid curve). The green line selects the rescaled energy $\epsilon = -2.5$ at $N=80$.
		(b)~Streamlines demonstrate the flow pattern of a sample point in phase space with low energy, $\epsilon = -2.5$ (see movie S3).
		(c)~The time-averaged Hamiltonian dynamics of $N=80$ vortices at $\epsilon = -2.5$ exhibits edge modes (see movie S4). Simulation time: $T = 8\times 10^3 T_0$.
		(d)~Mean-field theory in a bean-shaped domain predicts edge modes ($\beta = 15$).
		(e, f)~The mean-field theory prediction (solid black curve) agrees quantitatively with time-averaged Hamiltonian dynamics (green circles), and deviates significantly from the uniform vortex distribution (solid horizontal line).} 
	\label{fig_bean_mf}
\end{figure*}

\subsection{Topological interpretation of edge modes}

Operators of the form $\nabla^2 - m^2$ have a topological interpretation in certain contexts, including in the theory of chiral hydrodynamics~\cite{dasbiswas2018topological}. In our system, this operator appears in the linearization of the mean-field equations \eqref{MFE} after neglecting angular momentum ($\gamma=0$)

\begin{align}\label{MFE_linear}
    \left(\nabla^2 -\frac{\beta}{Z}\lambda^2 \right)\tilde{\psi} = -\frac{\lambda}{Z}
\end{align}
This linearization is valid when $|\beta\lambda\tilde{\psi}|$ is small, which holds close to the boundary, see Eq.~\eqref{MFE}. Although this equation has exponentially decaying solutions, the assumption that $|\beta\lambda\tilde{\psi}|$ is small means that these solutions are not valid everywhere. The true decay behavior is given by exact solutions to the full nonlinear equation, such as \eqref{mf_sol}. Setting $\psi_0 = \tilde{\psi} - 1/\beta\lambda$ removes the constant term in equation
\eqref{MFE_linear}
\begin{align}\label{MFE_linear_hom}
    \left(\nabla^2 -\beta_1\lambda^2 \right)\psi_0 = 0
\end{align}
where $\beta_1 = \beta/Z$ as before.
\par
The topological interpretation of this simplified equation comes from taking square root of the differential operator
\begin{align*}
M = \begin{pmatrix}
-i \partial_y && -\partial_x + \lambda\sqrt{\beta_1} \\
\partial_x + \lambda\sqrt{\beta_1} && i\partial_y
\end{pmatrix} , \qquad M^2 = -\left(\nabla^2 -\beta_1\lambda^2 \right)I_2
\end{align*}
where $I_2$ is the $2\times 2$ identity matrix. The matrix $M$ is the 2D Dirac Hamiltonian, and has topological properties~\cite{dasbiswas2018topological}. In the Fourier representation, $ -i\partial_x\mapsto q_x$ and $  -i\partial_y\mapsto q_y$, we can write the matrix operator $M$ as
\begin{align}
M(q_x,q_y,\lambda) = \begin{pmatrix}
q_y && -iq_x + \lambda\sqrt{\beta_1} \\
iq_x + \lambda\sqrt{\beta_1} && -q_y
\end{pmatrix}
\label{e:M_fourier}
\end{align}
When viewed as a function on the parameter space $(q_x,q_y,\lambda)$, a topological quantity known as a Chern number~\cite{dasbiswas2018topological,fruchart2013introduction} can be associated with $M$ (Appendix). This invariant measures the winding of the eigenvectors of $M$ across the $(q_x, q_y)$ plane, and depends only on the sign of $\lambda$. As a consequence, regions in the physical $(x,y)$ space where $\lambda$ changes sign are special, and host localized steady-state flows~\cite{dasbiswas2018topological} (Appendix). This corresponds exactly to our physical picture of vortex dynamics. Our systems of positive vortices have $\lambda=1$, and are surrounded by negative image vortices with $\lambda=-1$ (Fig.~\ref{fig_sphere_density}). At the boundary, where $\lambda$ changes sign, we observe statistical edge modes. This argument requires $\beta<0$, and so is not incompatible with the high energy regime, where edge modes are not observed. Although this description~\cite{dasbiswas2018topological,murugan2017topologically} is not the standard picture of topological edge modes~\cite{fruchart2013introduction}, it presents a possible avenue for exploring connections with other topological phenomena.

\subsection{Mean-field theory in nonconvex domains}

Low-energy edge modes also survive in nonconvex domains at large $N$. As in the disk, the rescaled energy $\epsilon$ defines low-energy states [Fig.~\ref{fig_bean_mf}(a)] which display statistical edge modes [Figs.~\ref{fig_bean_mf}(b)  and \ref{fig_bean_mf}(c)]. Expressed as a nonlinear eigenvalue problem, the mean-field equation without angular momentum is
\begin{align}\label{NLEVP}
\nabla^2 \phi = \beta_1 \lambda^2 e^\phi , \qquad\qquad \phi|_{\partial \Omega} = 0
\end{align}
where $\rho$ is obtained as
\begin{align*}
\rho = \frac{1}{\beta}\nabla^2 \phi ,\qquad\qquad \beta = \beta_1\lambda^2 \int_\Omega d^2\mathbf{x} \,  e^\phi
\end{align*}
The numerical solution of this equation agrees with time-averaged simulations of the Hamiltonian dynamics [Fig.~\ref{fig_bean_mf}(c)-\ref{fig_bean_mf}(f)]. Mean-field theory thus remains valid for irregular, nonconvex domains at low energy, and produces statistical states with edge modes. In particular, any smooth solution of (\ref{NLEVP}) with $\beta_1>0$ gives a vortex density which is maximized on the boundary. To see this, observe that $\phi$ is subharmonic, $\nabla^2 \phi > 0$. By the maximum principle for subharmonic functions~\cite{menon2018statistical}, $\phi$ must be maximized on the boundary of its domain. Since $\phi$ is constant on the boundary, we have 
\[
\phi(\mathbf{x}_b) \geq \phi(\mathbf{x}) \qquad \forall \mathbf{x}_b \in \partial\Omega, \; \mathbf{x}\in \Omega
\]
The vortex density $\rho$ is an increasing function of $\phi$,
\begin{align*}
\rho = \frac{1}{\beta} \nabla^2 \phi = \frac{\beta_1\lambda^2}{\beta} e^{\phi} 
\end{align*}
Thus $\rho$ is also maximized on the boundary.

\section{Conclusions}

By explicit treatment of boundaries, we have shown that the strongly interacting point vortex system displays statistical edge modes at low energy. These edge modes are robust to changes in geometry, surviving in convex and nonconvex domains, and persist in the large vortex number limit. An interesting future extension could be the investigation of similar phenomena in systems with multiple vortex species, and in more complicated domains, including multiply connected domains, for which there exist a wide variety of conformal mapping techniques~\cite{2005CrowdyMarshall,baddoo2019periodic,vasconcelos2015secondary,aref2007point}. Although our edge modes are not explicitly topological, there might exist links to topological edge modes in other systems, as the mean-field theory used here bears some resemblance to the description of Chern-Simons vortices~\cite{dunne1999aspects}. More generally, the above results raise the interesting question of whether an explicit treatment of boundaries could yield insights into the real-space mechanisms underlying other topological edge-mode  phenomena~\cite{delplace2017topological,van2016spatiotemporal, souslov2019topological,dasbiswas2018topological,1994Thouless,PhysRevB.25.2185}.

\section{Acknowledgements}

We are grateful to Henrik Ronellenfitsch, Peter Baddoo, Pearson Miller and Martin Zwierlein for valuable discussions and comments. This work was supported by a MathWorks Fellowship~(V.P.P.), and the Robert E. Collins Distinguished Scholar Fund of the MIT Mathematics Department (J.D.).

\section*{Appendix}

\subsection{Topological aspects of the mean-field equation}

\subsubsection{Chern numbers}

The linearized mean-field equation \eqref{MFE_linear_hom} can be interpreted topologically by virtue of its associated Dirac matrix operator from Eq.~\eqref{e:M_fourier}, 
\begin{align*}
M(q_x,q_y,\lambda) = \begin{pmatrix}
q_y && -iq_x + \lambda\sqrt{\beta_1} \\
iq_x + \lambda\sqrt{\beta_1} && -q_y
\end{pmatrix}
\end{align*}
Each eigenvector of $M$ corresponds to a Chern number. We sketch the calculation of these quantities following Refs. ~\cite{fruchart2013introduction, dasbiswas2018topological}. To this end, we first rewrite $M$ by formally expressing the matrix parameters $(q_x,q_y,\lambda\sqrt{\beta_1})$ in terms of spherical polar coordinates
\begin{align*}
    \lambda\sqrt{\beta_1} = h\cos\phi\sin\theta , \quad q_x = h\sin\phi\sin\theta , \quad q_y = h\cos\theta
\end{align*}
Then 
\begin{align*}
M = \begin{pmatrix}
h\cos\theta && he^{-i\phi}\sin\theta \\
he^{i\phi}\sin\theta && -h\cos\theta
\end{pmatrix}
\end{align*}
The eigenvectors of $M$ are
\begin{align*}
\Psi_- = \begin{pmatrix}
e^{-i\phi}\sin\theta/2\\
-\cos\theta/2
\end{pmatrix} , \qquad  
\Psi_+ = \begin{pmatrix}
e^{-i\phi}\cos\theta/2\\
\sin\theta/2
\end{pmatrix} 
\end{align*}
The Berry phase for each eigenvector is
\begin{align*}
A^-_{\theta} &= -i \langle \Psi_-| \partial_\theta \Psi_-\rangle = 0 ,\qquad A^-_{\phi} = -i \langle \Psi_-| \partial_\phi \Psi_-\rangle = -\sin^2\frac{\theta}{2}\\
A^+_{\theta} &= -i \langle \Psi_+| \partial_\theta \Psi_+\rangle = 0 ,\qquad A^+_{\phi} = -i \langle \Psi_+| \partial_\phi \Psi_+\rangle = -\cos^2\frac{\theta}{2}
\end{align*}
This gives the Berry curvature in polar coordinates
\begin{align*}
F_{\theta\phi}^\pm = \partial_{\theta} A_{\phi}^\pm - \partial_{\phi} A_{\theta}^\pm = \pm \frac{1}{2}\sin\theta
\end{align*}
Transforming back to Cartesian coordinates $(q_x,q_y,\lambda\sqrt{\beta_1})$, we find
\begin{align*}
    F^\pm = \pm \frac{1}{2}\frac{\lambda\sqrt{\beta_1}}{q_x^2 + q_y^2 + \lambda^2\beta_1 }
\end{align*}
Finally, the Chern number for each eigenvector is given by integrating $F$ over $q_x$ and $q_y$
\begin{align*}
    C^\pm = \int \frac{dq_x\,dq_y}{2\pi} F^\pm = \pm \frac{1}{2}\,\text{sgn}(\lambda)
\end{align*}
Regions of the $(x,y)$ plane with $\lambda$'s of different signs are therefore associated with different Chern numbers. The nonzero difference in Chern number across such a boundary is responsible for presence of a localized edge mode
\begin{align*}
    |\Delta C| = |C^\pm(\lambda>0) - C^\pm (\lambda<0)| = 1
\end{align*}

\subsubsection{Localized states}

Another way to see that localized states occur when $\lambda$ changes sign is to look for solutions to the following equation~\cite{fruchart2013introduction}
\begin{align*}
M\begin{pmatrix}
f\\
g\end{pmatrix} = 
\begin{pmatrix}
-i \partial_y && -\partial_x + \lambda\sqrt{\beta_1} \\
\partial_x + \lambda\sqrt{\beta_1} && i\partial_y
\end{pmatrix}
\begin{pmatrix}
f\\
g\end{pmatrix} = 0
\end{align*}
for some functions $f$ and $g$. Assume that there is a boundary at $x=0$, with positive vortices in the region $x>0$ and negative (image) vortices in the region $x<0$. For ease of notation, set $m = \lambda\sqrt{\beta_1}$, so $m>0$ when $x>0$ and $m<0$ and $x<0$. To solve these equations, we assume that $m$ is continuous but changes rapidly from $-\sqrt{\beta_1}$ to +$\sqrt{\beta_1}$ across the boundary at $x=0$. We will look for separable functions $f$ and $g$, which decay for large $x,y$
\begin{align*}
-if_y -g_x + m(x)g &= 0\\
f_x + m(x) f + i g_y &=0
\end{align*}
where subscripts denote derivatives here. Consider the first equation
\begin{align*}
-if_y - \left(\partial_x - m(x)\right)g = 0
\end{align*}
Since $m(x)$ and $x$ have the same sign, $g$ will grow exponentially. Therefore a bounded solution must have $g=0$. This gives $f_y=0$ immediately, so the only remaining equation is
\begin{align*}
f_x = -m(x) f
\end{align*}
This has solution
\begin{align*}
f(x) \propto \exp{\left(-\int_0^x dx'\,m(x')\right)}
\end{align*}
As expected, this solution is sharply peaked around the boundary $x=0$, and decays rapidly away from $x=0$.
\par
While the linearized mean-field equation~\eqref{MFE_linear_hom} enables an interpretation within the conventional linear-operator framework of  topologically protected modes, the more accurate description of the strongly interacting vortex system is provided by its nonlinear mean-field equation~\eqref{MFE}.

\subsection{Equilibration at large $N$}

\begin{figure*}[ht]
	\centering
	\includegraphics[width=\columnwidth]{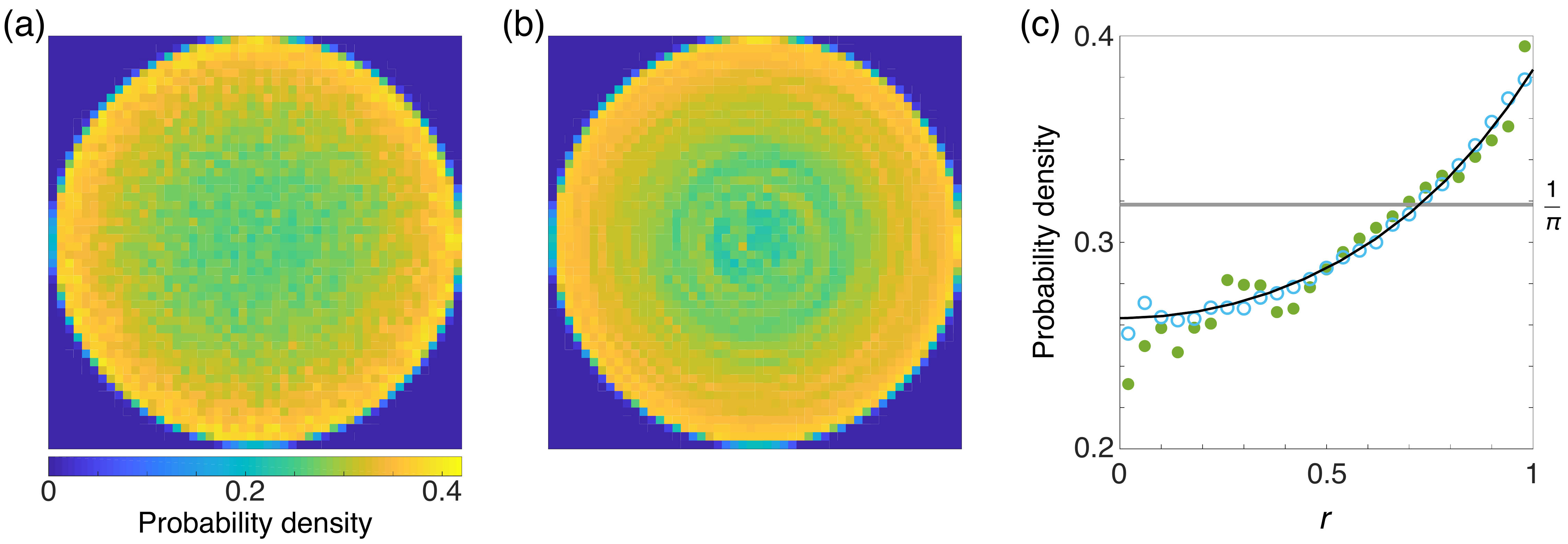}
	\caption{Hamiltonian dynamics, Monte Carlo sampling, and mean-field predictions are in quantitative agreement at low energy ($N=500$).
		(a)~Monte Carlo sampling at fixed energy ($\epsilon = -2.5$) and varying angular momentum reveals a low energy edge mode ($n = 3540$ samples of $N=500$ vortices).
		(b)~The time-averaged Hamiltonian dynamics of $N=500$ vortices at $\epsilon = -2.5$ exhibits edge modes. Simulation time: $T = 2\times 10^3 T_0$.
		(c)~Mean-field theory in the disk, at zero angular momentum, predicts edge modes ($\beta = 5.2, \gamma=0$). Away from $r=0$, quantitative agreement is observed between the mean-field theory prediction (solid black curve), time-averaged Hamiltonian dynamics (filled green circles) and Monte Carlo sampling (empty blue circles). Solid horizontal line shows uniform vortex distribution.} 
	\label{fig_N500}
\end{figure*}

In the systems of positive point vortices considered here, time-averaged dynamical simulations are found to agree with vortex distributions obtained by Monte Carlo sampling at fixed energy, even for large $(N=500)$ vortex numbers (Fig.~\ref{fig_N500}). Although there are mathematical questions which remain unresolved, it is believed that this observed ergodicity is an appropriate assumption for the point vortex model under certain conditions~\cite{esler2017equilibrium,chavanis2012kinetic}. In other cases, the system is known to be non-ergodic~\cite{lim1990existence,khanin1982quasi}. Numerical studies indicate evidence for ergodicity at intermediate energies for mixed vortex systems (with $\lambda_a = \pm 1$)  in bounded~\cite{esler2015universal,esler2017statistics} and periodic~\cite{esler2017statistics} domains.  On the sphere, the ergodic hypothesis appears to hold for a range of energies~\cite{dritschel2015ergodicity}. At very low energies, however, dipole formation in mixed vortex systems can produce quasi-stationary states~\cite{venaille2015violent,esler2017statistics}. Furthermore, it is possible that the relaxation time scales unfavorably with the vortex number $N$ in certain scenarios~\cite{chavanis2012kinetic}. Since we only consider positive vortices, we do not expect dipole formation to obstruct equilibration at very low energies. However, relaxation timescale issues could be responsible for the discrepancy we observe in the vortex distributions obtained from dynamical simulations, Monte Carlo sampling and mean-field theory, near the center ($r\to 0$) of the disk [Fig.~\ref{fig_N500}(c)].

The edge mode obtained for $N=500$ vortices at rescaled energy $\epsilon = -2.5$ (Fig.~\ref{fig_N500}) is shallower than that obtained for $(N,\epsilon) = (80,-2.5)$ (Figs.~\ref{fig_disk_mf} and \ref{fig_bean_mf}). In particular, the mean-field solution for the $(N,\epsilon) = (80,-2.5)$ edge mode has $\beta=15$, wheras the $(N,\epsilon) = (500,-2.5)$ edge mode has $\beta=5.2$. However, in both cases, the mean-field solution \eqref{mf_sol} accurately describes the vortex density
\begin{align*}
\rho(r) = \frac{8}{8\pi+\beta} \frac{1}{\left(1 - \frac{\beta r^2}{8\pi+\beta}\right)^2} , \qquad \beta > -8\pi
\end{align*}
This suggests that, by choosing an $\epsilon$ which corresponds to a larger $\beta$, a sharper edge mode can be found for $N=500$.

\subsection{Mean-field theory from the canonical ensemble}

The mean-field equations \eqref{MFE} can be derived using the canonical ensemble. Following the presentation in Ref.~\cite{Chavanis2014}, the canonical partition function for $N$ vortices in a domain $\Omega$ is
\begin{align*}
Z(\beta) = \int_{\Omega^N} d\xi\,e^{-\beta\mathcal{H}(\xi)} 
\end{align*}
where $\xi = (x_1, y_1, ... , x_N, y_N)$ is a point in $\Omega^N$. This can be rewritten in terms of the dimensionless vortex density $\rho = \omega T_0$. Let $\Gamma[\rho]$ be the functional corresponding to the total vortex density
\begin{align*}
\Gamma[\rho] = \int_{\Omega} d^2 \mathbf{x} \, \rho(\mathbf{x}) = 1
\end{align*}
Setting $W[\rho]$ to be the number of microstates $\{\xi\}$ corresponding to the macrostate $\{\rho(\mathbf{x})\}$, allows us to write $Z(\beta)$ as an integral over fields $\rho(\mathbf{x})$
\begin{align*}
Z(\beta) = \int_{\Omega^N} d\xi\,e^{-\beta\mathcal{H}(\xi)} = \int \mathcal{D}\rho \, W[\rho] \,\delta\left(\Gamma[\rho] - 1\right) e^{-\beta\mathcal{H}[\rho]}
\end{align*}
The energy is
\begin{align*}
\mathcal{H}[\rho] = -T_0^2\int_{\Omega} d^2\mathbf{x}\,d^2\mathbf{x}'\, \rho(\mathbf{x}) G(\mathbf{x} - \mathbf{x}') \rho(\mathbf{x}')  =  T_0^2 \int_{\Omega} d^2\mathbf{x}\, \tilde{\psi}\rho
\end{align*}
where the second equality uses $\tilde{\psi} = \psi T_0$ from \eqref{MFT_functional}.
By a counting argument~\cite{menon2018statistical,Chavanis2014}, the entropy $S[\rho]$ is related to $W[\rho]$ by
\begin{align*}
S[\rho] = -\int_{\Omega} d^2\mathbf{x}\, \rho \log\rho = \log W[\rho]
\end{align*}
The partition function $Z(\beta)$, and the probability $P[\rho]$ of the density $\rho$ are therefore
\begin{align*}
Z(\beta) &= \int \mathcal{D}\rho \, \delta\left(\Gamma[\rho] - 1\right) e^{S[\rho]-\beta\mathcal{H}[\rho]}\\
P[\rho] &\propto \delta\left(\Gamma[\rho] - 1\right) e^{S[\rho]-\beta\mathcal{H}[\rho]}
\end{align*}
The most probable distribution follows from maximizing $S - \beta\mathcal{H}$ subject to $\Gamma[\rho]=1$. In other words, we must maximize the functional
\begin{align*}
\mathcal{I}[\rho] = S - \beta\mathcal{H} - \alpha \int_{\Omega} d^2\mathbf{x} \, \rho
\end{align*}
Upon rescaling $\beta$, this is exactly the maximization in \eqref{MFT_functional} without angular momentum ($\gamma=0$). As before, this maximization yields the mean-field equations \eqref{MFE} with $\gamma=0$.

\bibliography{references}

\end{document}